\documentclass[format=sigconf,anonymous=false,review=false,screen=true]{acmart}

\AtBeginDocument{%
  \providecommand\BibTeX{{%
    \normalfont B\kern-0.5em{\scshape i\kern-0.25em b}\kern-0.8em\TeX}}}

\copyrightyear{2023}
\acmYear{2023}
\setcopyright{acmlicensed}\acmConference[KDD '23]{Proceedings of the 29th ACM SIGKDD Conference on Knowledge Discovery and Data Mining}{August 6--10, 2023}{Long Beach, CA, USA}
\acmBooktitle{Proceedings of the 29th ACM SIGKDD Conference on Knowledge Discovery and Data Mining (KDD '23), August 6--10, 2023, Long Beach, CA, USA}
\acmPrice{15.00}
\acmDOI{10.1145/3580305.3599900}
\acmISBN{979-8-4007-0103-0/23/08}

\makeatletter
\def\blfootnote{\gdef\@thefnmark{}\@footnotetext}
\makeatother

\usepackage{graphicx}
\usepackage[caption=false]{subfig}
\usepackage{color}
\usepackage{algorithm}
\usepackage{algorithmic}

\usepackage{microtype}
\usepackage{amsmath,amsfonts}
\usepackage{mathtools}
\usepackage{mathrsfs}
\usepackage{bm}
\usepackage{booktabs, multirow}
\usepackage{makecell}
\usepackage{stfloats}
\usepackage{xcolor}
\usepackage{arydshln}
\usepackage{pifont}
\usepackage{tikz}
\newcommand*{\circled}[1]{\lower.7ex\hbox{\tikz\draw (0pt, 0pt)%
    circle (.5em) node {\makebox[1em][c]{\small #1}};}}

\begin{document}

\title{RLTP: Reinforcement Learning to Pace for Delayed Impression Modeling in Preloaded Ads}

\author{Penghui Wei$^{ \dagger}$}
\affiliation{%
  \institution{Alibaba Group}
  \city{} 
  \country{} 
}
\email{wph242967@alibaba-inc.com}

\author{Yongqiang Chen$^{ \dagger}$}
\affiliation{%
  \institution{Alibaba Group}
  \city{} 
  \country{} 
}
\email{shiwei.cyq@alibaba-inc.com}

\author{Shaoguo Liu}
\affiliation{%
  \institution{Alibaba Group}
  \city{} 
  \country{} 
}
\email{shaoguo.lsg@alibaba-inc.com}

\author{Liang Wang}
\affiliation{%
  \institution{Alibaba Group}
  \city{} 
  \country{} 
}
\email{liangbo.wl@alibaba-inc.com}

\author{Bo Zheng}
\affiliation{%
  \institution{Alibaba Group}
  \city{} 
  \country{} 
}
\email{bozheng@alibaba-inc.com}

\thanks{$^{\dagger}$Co-first authorship. }

\renewcommand{\authors}{Penghui Wei, Yongqiang Chen, Shaoguo Liu, Liang Wang and Bo Zheng}
\renewcommand{\shortauthors}{Penghui Wei, Yongqiang Chen, Shaoguo Liu, Liang Wang, \& Bo Zheng}

\begin{abstract}
To increase brand awareness, many advertisers conclude contracts with advertising platforms to purchase traffic and  deliver advertisements to target audiences. In a whole delivery period, advertisers desire a certain \textbf{impression count} for the ads, and they expect that the \textbf{delivery performance} is as good as possible. Advertising platforms employ real-time pacing algorithms to satisfy the demands. However, the delivery procedure is also affected by publishers. Preloading is a widely used strategy for many types of ads (e.g., video ads) to make sure that the response time for displaying is legitimate, which results in \textit{delayed} impression phenomenon. 

In this paper, we focus on a new research problem of impression pacing for preloaded ads, and propose a \textbf{R}einforcement \textbf{L}earning \textbf{T}o \textbf{P}ace framework \textsf{RLTP}. It learns a pacing agent that sequentially produces selection probabilities in the whole delivery period. To jointly optimize the objectives of impression count and delivery performance, \textsf{RLTP} employs tailored reward estimator to satisfy guaranteed impression count, penalize over-delivery and maximize traffic value. Experiments on large-scale datasets verify that \textsf{RLTP} outperforms baselines by a large margin. We have deployed it online to our advertising platform, and it achieves significant uplift to core metrics including delivery completion rate and click-through rate.

\end{abstract}

%
%


\maketitle

\section{Introduction}

\begin{figure}[t]
\centering
\centerline{\includegraphics[width=\columnwidth]{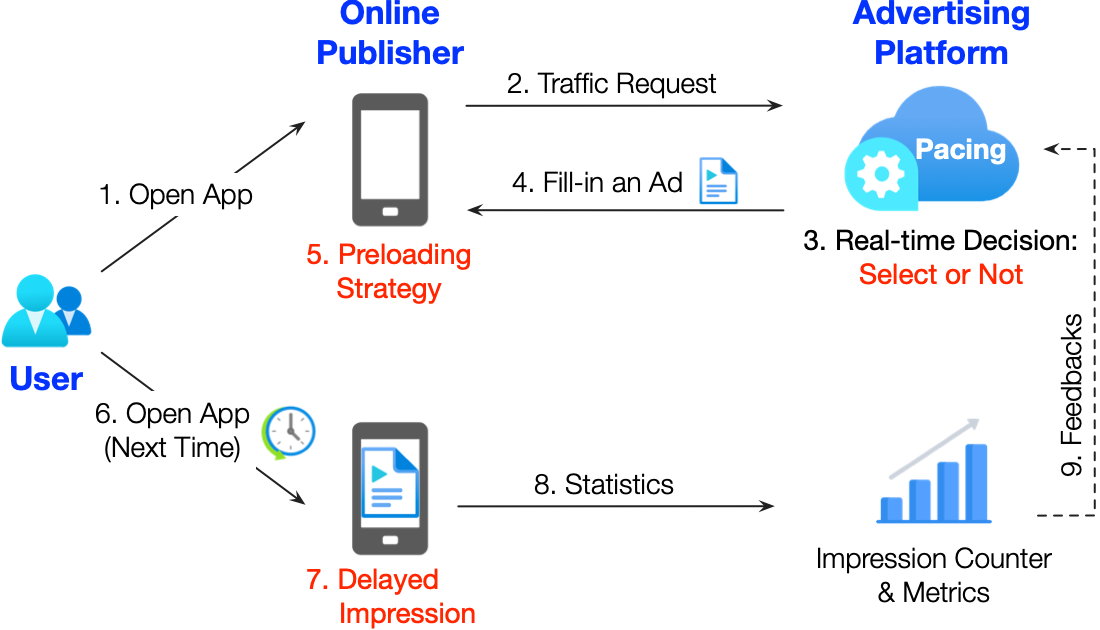}}
\caption{The overall delivery procedure. Due to the \textit{preloading} strategy from the publisher, the ad filled by the advertising platform cannot be immediately displayed at current request. We call this phenomenon as \textit{delayed impression}.}
\label{fig:delivery}
\end{figure}

Merchants usually achieve their customized marketing goals via delivering advertisements on online publishers. To increase brand awareness, many advertisers conclude contracts with advertising platforms to purchase traffic, and deliver advertisements to target audiences. Figure~\ref{fig:delivery} gives an illustration of delivery procedure for brand advertising. When a user opens an App and browses a specific ad exposure position (i.e., sends an ad request to the publisher and ad exchange), the advertising platform estimates the traffic value and determines \textit{whether to select} this request. If it makes a selection decision, it will fill-in an ad -- which is expected to be displayed to the user -- to the online publisher. Finally, the publisher determines a final ad that will be displayed based on its own strategies, such as preloading and frequency controlling. Therefore, at current request, it is possible that the ad filled by the platform is not displayed, but a preloaded ad from another advertising platform is displayed.

For system performance and user experience, \textit{preloading} is a widely used strategy for many types of ads (e.g., video ads) to ensure that the response time for displaying is legitimate. If  publisher employs preloading, the filled ad from advertising platform may be displayed at \textit{one of the next requests} sent by the same user (e.g., the next time the user opens the App, as in Figure~\ref{fig:delivery}), or even may not be displayed if the user does not send any requests after this time. 
Therefore, \textit{delayed impressions} of ads is a typical phenomenon in the ad delivery procedure that exists preloading strategy.

To ensure the effectiveness of ad delivery, in a whole delivery period (e.g., one day), advertisers usually have two types of demands: 1) The first is about \textbf{impression count}. Advertisers desire a certain impression count for an ad to reach target audiences. Advertising platforms need to guarantee that the actual impression count is as close as possible to this desired value, and also avoid over-delivery which means that the actual impression count is much larger than the guaranteed one. Besides, it is also desirable that all impressions are evenly distributed to all time windows of a period, in other words we seek the \textbf{smoothness} of the overall delivery. The advantage is that the ads can reach active audiences at different time periods. 2) The second demand is about \textbf{delivery performance}. Advertisers also expect that the delivery performance, evaluated by click-through rate (CTR) and post-click metrics like conversion rate, is as good as possible. This demand reflects that advertisers have a strong desire to obtain high-value traffic. 

From the perspective of advertising platforms, the delayed impression phenomenon brings challenges to optimize the above objectives. If the delivery procedure is \textit{not} affected by publishers' strategies (i.e., after the advertising platform makes a selection decision to a request, the filled ad will be \textit{immediately} displayed to the user), the above demands can be satisfied by traditional {pacing algorithms}~\cite{bhalgat2012online,agarwal2014budget,xu2015smart}: we can adjust the {selection probability} at each time window (e.g., 5 minutes) based on  the difference between actual impressions and expected impressions to satisfy impression count demand, and perform grouped adjusting~\cite{xu2015smart} based on predicted CTR to satisfy performance demand.  However, with preloading strategy, advertising platforms can only determine the \textit{selection} of requests but the final \textit{impressions} are controlled by publishers. Therefore, without immediate feedbacks, it is challenging for platforms to guarantee advertiser demands.

In this paper, we focus on a new research problem of impression pacing for preloaded ads, aiming to guarantee that both impression count and delivery performance demands are satisfied. Formally, we formulate this problem as a sequential decision process. At each time window in a delivery period, the pacing algorithm produces a {selection probability} that determines whether to select a given traffic request in this window. The current selection probability should be affected by previous ones and also affects next ones, because the utility of the produced sequential decisions is evaluated after the whole delivery period is terminated. 

The problem of impression pacing for preloaded ads has two characteristics. 1) \textit{The feedback signal is typically delayed} due to the delayed impression phenomenon, i.e., we observe whether the demands are satisfied until the end point of whole delivery period. 2) Besides, \textit{it is hard to exactly mimic or model the delay situation}, because the publishers may change their strategy and advertising platforms are unaware of this.  Such two characteristics inspire us to employ reinforcement learning, where the  policy 1) is encouraged to obtain \textit{long-term cumulative rewards} other than immediate ones, and 2) is learned via a \textit{trial-and-error interaction process}, without the need of estimating environment dynamics. 

Motivated by the above considerations, we propose \textsf{RLTP}, a \textbf{R}einforcement \textbf{L}earning \textbf{T}o \textbf{P}ace framework. In \textsf{RLTP}, the combination of user and online publisher is regarded as the environment that provides states and rewards to train the pacing agent that sequentially produces selection probability at each decision step. To jointly optimize the objectives of impression count and delivery performance, the reward estimator in \textsf{RLTP} considers  key aspects to guide the agent policy learning via satisfying the guaranteed impression count, maximizing the traffic value, and penalizing the over-delivery as well as the drastic change of selection probability. We conduct offline and online experiments on large-scale industrial datasets, and results verify that \textsf{RLTP} significantly outperforms well-designed pacing algorithms for preloaded ads.

 The contributions of this paper are: 
\begin{itemize}
    \item To the best of our knowledge, this work is the first attempt that focuses on impression pacing algorithms for preloaded ads, which is very challenging for advertising platforms due to the delayed impression phenomenon. 
    \item To jointly optimize both objectives (impression count and delivery performance) for preloaded ads, we propose a reinforcement learning to pace framework \textsf{RLTP} for delayed impression modeling. We design tailored reward estimator to satisfy impression count, penalize over-delivery and maximize traffic value. 
    \item Experiments on large-scale industrial datasets show that \textsf{RLTP} consistently outperforms compared algorithms. It has been deployed online in a large advertising platform, and  results verify that  \textsf{RLTP} achieves significant uplift to delivery completion rate and click-through rate.
\end{itemize}

\section{Preliminaries}
We start from the formulation of impression pacing problem for preloaded ads. We then introduce a vanilla solution based on PID controller to tackle the problem, and show its limitations.

\subsection{Problem Formulation}

\subsubsection{\textbf{Objective}}\label{sec:formulation_objective}

For a whole delivery period, let $N_\mathrm{target}$ denote the desired impression count (set by the advertiser). We assume that the number of traffic requests $N_\mathrm{req}$ from publisher is sufficient:
\begin{equation}
    N_\mathrm{req} \gg N_\mathrm{target}\,, 
\end{equation}
however the value of $N_\mathrm{req}$ is unknown before delivery. During ad delivery, advertisers have two types of demands that advertising platforms need to guarantee:  
\begin{itemize}
    \item \textbf{Impression count}: for a whole delivery period, the total \textit{actual impression count} $N_\mathrm{imp}$ meets the condition of $N_\mathrm{imp}\leq N_\mathrm{target}+\epsilon$, the closer the better:
    \begin{equation}\label{eq:impression}
        \begin{aligned}
        \min \quad & N_\mathrm{target} -  N_\mathrm{imp}  & \scriptstyle{\text{; The closer the better.}} \\
        \textrm{s.t.} \quad & N_\mathrm{imp} - N_\mathrm{target} \leq \epsilon  & \scriptstyle{\text{; Avoid over-delivery.}} \\
          &\epsilon\geq0    \\
        \end{aligned}
    \end{equation}
    where $\epsilon\geq0$ is a tolerance value of over-delivery, and usually very small. Serious over-delivery (i.e., $N_\mathrm{imp}> N_\mathrm{target}+\epsilon$) is unacceptable, which results in reimbursement. Besides, the delivery process should be \textbf{smoothness}, which means that all impressions are desirable to be as evenly as possible distributed over the time windows of the period. 
    \item \textbf{Delivery performance}: In this work, we employ CTR as the metric for delivery performance. We aim to maximize the overall performance and ensure that the ads are displayed to high-value traffic. 
\end{itemize}

\begin{figure}[t]
\centering
\centerline{\includegraphics[width=\columnwidth]{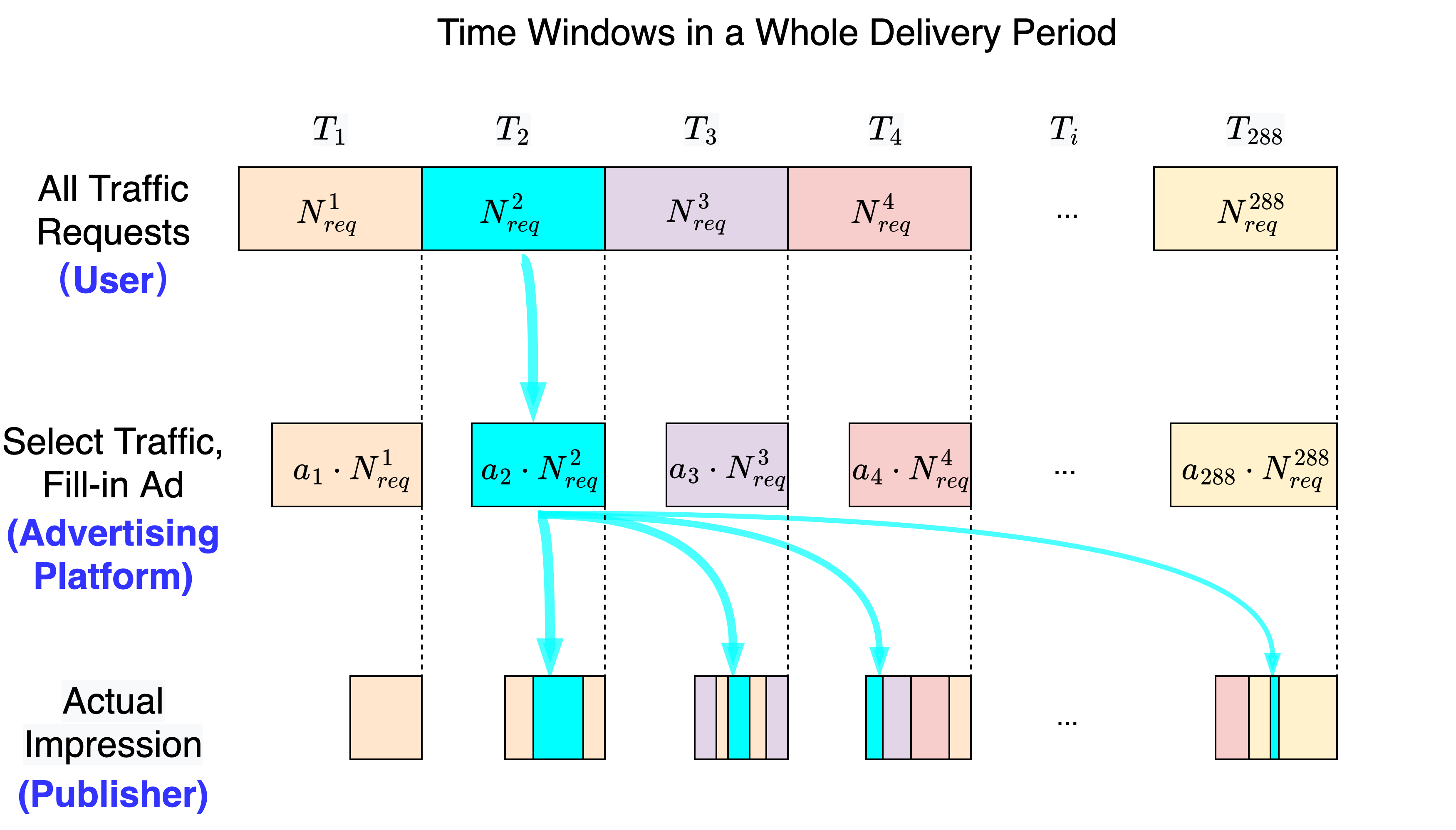}}
\caption{(Better viewed in color) Illustration of delayed impression phenomenon under the preloading strategy controlled by publisher. At each time window $T_i$, the actual impression ads may come from the filled ads at current $T_i$ and previous windows $\left\{T_1,T_2,\ldots,T_{i-1}\right\}$.}
\label{fig:delayedmodeling}
\end{figure}

\subsubsection{\textbf{Delayed Impression in Preloaded Ads}}
Due to the publisher's preloading strategy, although the advertising platform determines to select a given traffic and then fills in an ad, its impression time is usually at \textit{one of the next requests} sent by the same user, or it even may not be displayed if the user does not send any requests after this time in the whole delivery period. 
We call this phenomenon as \textit{delayed impression}.

Figure~\ref{fig:delayedmodeling} gives an  illustration of delayed impression in a whole delivery period. Consider that the whole delivery period is divided into a number of equal-length time windows $\{T_1,T_2,\ldots,T_L\}$, where the interval length of each window is denoted as $\Delta T$. At each time window $T_i$'s start point, the \textbf{pacing algorithm} of our advertising platform need to produce a \textit{selection probability} $a_i\in [0,1]$ that determines whether to select a given traffic request in this time window, and the goal is to satisfy advertiser demands of impression count and delivery performance.

Consider the $i$-th time window $T_i$, let $N_\mathrm{req}^{i}$ denote the number of {total requests} sent from users, and thus $a_i\cdot N_\mathrm{req}^{i}$ requests are {selected} by our advertising platform and then they will be filled in ads. However, under the preloading strategy from online publisher, the \textit{actual impression count} of these filled ads can only be observed at the end point of the whole period due to the delayed impression phenomenon. The feedback signal  for learning pacing algorithm is typically delayed.

Formally, let  $N_\mathrm{imp}^{i}$ denote the actual impression count of the selected ads in time window $T_i$. As shown in Figure~\ref{fig:delayedmodeling}, these impressions are scattered over the time windows $\{T_i, T_{i+1},\ldots, T_L\}$:
\begin{equation}
    N_\mathrm{imp}^{i} = \beta_i N_\mathrm{imp}^{i} + \beta_{i+1} N_\mathrm{imp}^{i} + \ldots \beta_{L} N_\mathrm{imp}^{i}\,,\mathrm{where }\sum_{j=i}^L \beta_j=1
\end{equation}
here we use delayed factor $\beta_j$ to denote the proportion between the impression count in time window $T_j$ and total impression count $N_\mathrm{imp}^{i}$. At the end of time window $T_i$, we can only \textit{observe} impressions $ \beta_i N_\mathrm{imp}^{i}$. Online publishers control the delayed factors and change them according to traffic dynamics and marketing goals. 

From the perspective of advertising platforms, it is hard to exactly mimic or model the delay situation (i.e., delayed factors), because they are unaware of such change from publishers in real-time. Therefore, it is very challenging for advertising platforms to satisfy both impression count and delivery performance demands for preloaded ads.

\subsubsection{\textbf{Impression Pacing for Preloaded Ads}}\label{sec:formulation}

We formulate the impression pacing problem for preloaded ads as a Markov decision process (MDP), which can be represented as a five-tuple $\left(\mathcal S, \mathcal A, \mathcal P, \mathcal R, \gamma \right)$:
\begin{itemize}
    \item $\mathcal S$ is the state space. For the time window $T_i$, the state $\boldsymbol s_i$ reflects the delivery status after windows $\left\{T_1,\ldots,T_{i-1}\right\}$. 
    \item $\mathcal A$ is the action space. The action $a_i$ at time window $T_i$ reflects the selection probability. 
    \item $\mathcal P: \mathcal S\times \mathcal A\times \mathcal S\rightarrow [0,1]$ is the transition function that describes the probability $p(\boldsymbol s'\mid \boldsymbol s, a)$ of transforming from state $\boldsymbol{s}$ to another state $\boldsymbol s'$ after taking the action $a$. 
    \item $\mathcal R: \mathcal S\times \mathcal A\rightarrow \mathbb R$ is the reward function, where $r(\boldsymbol{s}, a)$ denotes the reward value of taking the action $a$ at the state $\boldsymbol{s}$. To achieve the delivery objective of advertisers, the reward $r()$ should reflect the delivery completeness, smoothness and performance. 
    \item $\gamma\in [0,1]$ is the discount factor that balances the importance of immediate reward and future reward.   
\end{itemize}
Note that we can also define a cost function $\mathcal C: \mathcal S\times \mathcal A\rightarrow \mathbb R$ that penalizes over-delivery (the constraint in Equation~\ref{eq:impression}), and re-formulate the problem as a constrained Markov decision process~\cite{altman1999constrained}. For simplicity, we employ Lagrange function that transforms it to an unconstrained form.

Given the above MDP formulation of impression pacing problem for preloaded ads, the optimization objective is to learn an \textit{impression pacing policy} $\pi: \mathcal S\rightarrow\mathcal A$ via maximizing the expected cumulative reward:
\begin{equation}
    \max_{\pi}\mathbb E\left[ \sum_{i=1}^L r(\boldsymbol{s}_i, a_i) \right]
\end{equation}
The key is to design reward estimator which satisfies both impression count and delivery performance demands (Section~\ref{sec:formulation_objective}). 

Without loss of generality, in our work let 1-day or 1-week be the whole delivery period, and 5-minutes be the interval $\Delta T$ of each time window. 
In the situation of 1-day delivery period, the pacing policy need to produce sequential selection probabilities at 0:00, 0:05, ..., 23:50 and 23:55, containing $L=288$ decisions. 

\subsection{Vanilla Solution and its Limitations}\label{baseline:pid}
Traditional {pacing algorithms} in online advertising ~\cite{abrams2007optimal,borgs2007dynamics,bhalgat2012online,agarwal2014budget,xu2015smart,geyik2020impression} focus on the budget pacing problem, aiming to spend budget smoothly over time for reaching a range of target audiences, and various algorithms are proposed. However they cannot be directly adopted to our problem, because in their algorithms the feedback signals used for controlling are \textit{immediate}, while in our problem the impression signals can only be observed at the end of the delivery period due to the preloading strategy. To our knowledge, there is no existing studies that can tackle the impression pacing problem with delayed impressions of preloaded ads.

\subsubsection{\textbf{Vanilla Solution: Prediction-then-Adjusting}}
To tackle the delayed impression problem, we introduce a vanilla pacing algorithm modified from~\cite{cheng2022adaptive}. It contains two stages: 1) prediction stage, which employs a deep neural network (DNN) to predict actual impression count at current time window, and  2) adjusting stage, which employs a proportional–integral–derivative (PID) controllerr~\cite{bennett1993development} to adjust selection probability based on the difference of desired impression count and predicted impression count. This is the current version of our production system's pacing policy. 

The motivation is  to estimate ``immediate'' feedbacks for controlling current delivery via employing historial delivery data. Specifically, at time window $T_i$'s start point, the first stage uses historial delivery data to learn a prediction model that mapping \textit{observed} impression count to \textit{actual} impression count $\hat N_\mathrm{imp}^{i}$ for each time window. The estimated actual values are used as ``immediate'' feedbacks for further controlling. The model input contains statistical information from previous windows such as observed impression count, completion rate and averaged selection probability. An instantiation of the prediction model is a deep neural network~\cite{covington2016deep}. 

Then the second stage is a PID controller, which adjusts the initial selection probability at each time window based on the difference between the estimated actual impression rate $\hat N_\mathrm{imp}^{i}/N_\mathrm{target}$ and expected impression rate $\widetilde N_\mathrm{target}^{i}/N_\mathrm{target}$ to sastify impression count demand. Here, the expected count $\widetilde N_\mathrm{target}^{i}$ for each window is given in advance based on prior knowledge, such as the desired delivery speed distribution over the current delivery, and previous statistics from historical delivery logs. Formally, the adjusted selection probability $a_i'$ is computed as:
\begin{equation}
\small
\begin{aligned}
    \mathrm{error}_i &= \widetilde N_\mathrm{target}^{i}/N_\mathrm{target} - \hat N_\mathrm{imp}^{i}/N_\mathrm{target}\\
    \mathrm{pid\_score} &= K_\mathrm{P}\cdot \mathrm{error}_i + K_\mathrm{I}\cdot \sum_{j=1}^i\mathrm{error}_j + K_\mathrm{D}\cdot \left(\mathrm{error}_i - \mathrm{error}_{i-1} \right) \\
    a_i' &= a_i\cdot \mathrm{pid\_score}
\end{aligned}
\end{equation}
where $K_\mathrm{P}$, $K_\mathrm{I}$ and $K_\mathrm{D}$ are non-negative hyperparameters. 

To further satisfy delivery performance demand, inspired by~\cite{xu2015smart}, the PID controller maintains multiple groups of $\{K_\mathrm{P},K_\mathrm{I},K_\mathrm{D}\}$, where each group represents different performance levels. Each request falls into a specific group based on a CTR prediction model. The intuition is that if a request has higher predicted CTR, the adjusted selection probability should also be higher.

\subsubsection{\textbf{Limitations}}
The above  solution has advantages such as simple and easy-to-deployment. It also has unavoidable limitations:

1) The first stage estimates ``immediate'' feedbacks for current delivery based on historial processes, which inevitably brings estimation bias because publishers may change delayed strategy.

2) The second stage requires the expected counts of all time windows for controlling. The values are static, and computed based on the statistics from historical data. However, the timeliness is weak because the static values are unaware of real-time responses from users and publishers during the current delivery period. Besides, there are many hyperparameters that are hard to set.
    
3) Finally, the two stages may not adapt to each other very well, because they work independently and thus it is hard to obtain optimal pacing performance.

\begin{figure}[t]
\centering
\centerline{\includegraphics[width=0.9\columnwidth]{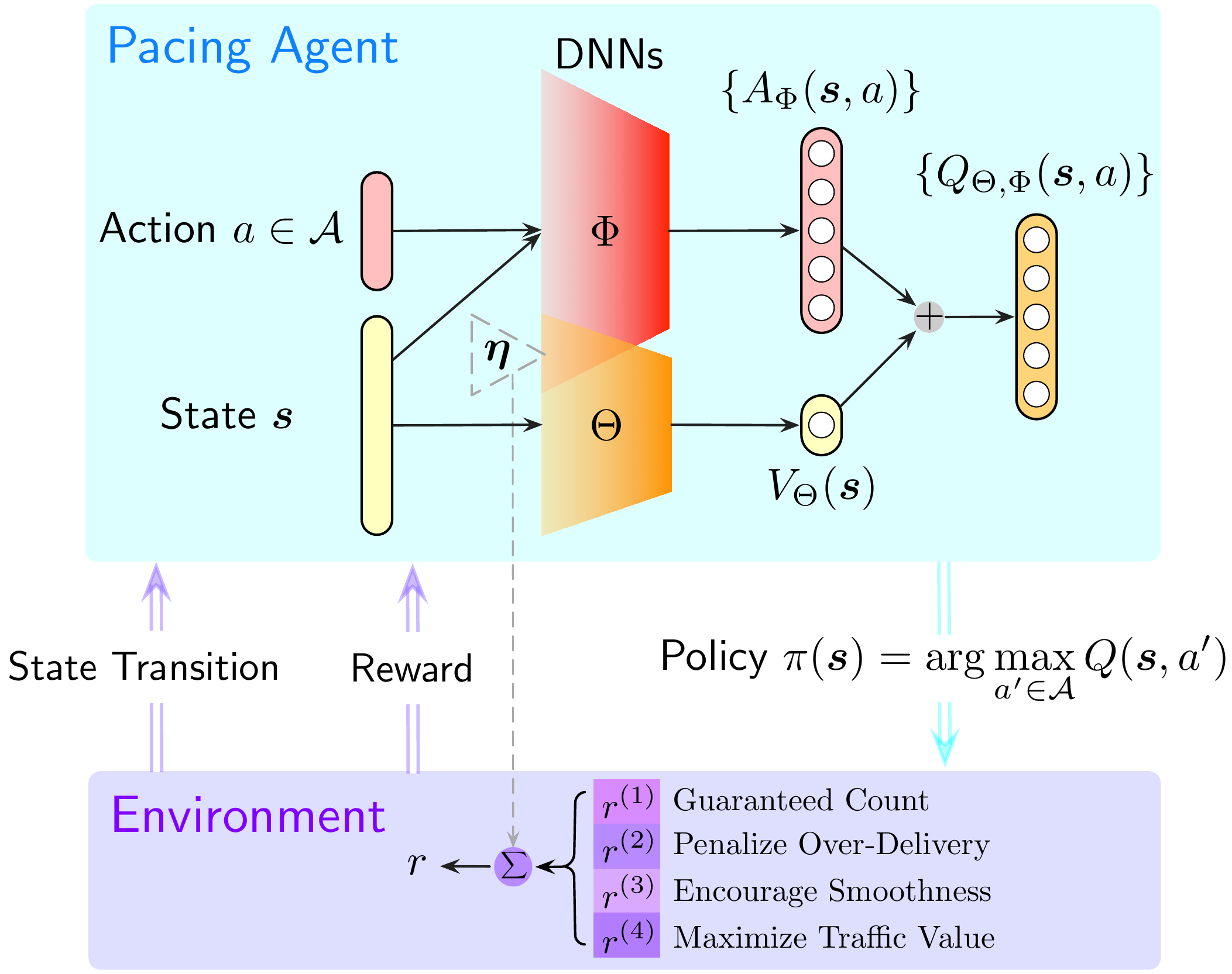}}
\caption{Overview of our reinforcement learning to pace framework \textsf{RLTP} for impression pacing on preloaded ads.}
\label{overview:rltp}
\end{figure}

\section{Proposed Approach: Reinforcement Learning to Pace}\label{sec:method}
During the ad delivery procedure, to handle the delayed impression phenomenon  under preloading strategy for satisfying advertiser demands, we employ reinforcement learning (RL) as the paradigm of our pacing algorithm. The advantage is that the  pacing policy is learned  through a trial-and-error interaction process guided by elaborate reward estimator and optimizes long-term value, which tackles the delayed feedback issue and does not need to exactly mimic the delay situation.

\subsection{Overview of Pacing Agent}
We propose a \textbf{R}einforcement \textbf{L}earning \textbf{T}o \textbf{P}ace framework \textsf{RLTP}, which learns a pacing agent that sequentially produces selection probability at each decision step. The ad delivery system is regarded as the environment, and the policy of \textsf{RLTP} is the agent. At each episode of $L$-step decisions for time windows $\{T_1,T_2,\ldots,T_L\}$, let $\tau =\left(\boldsymbol{s}_1,a_1,r_1,\ldots,\boldsymbol{s}_L,a_L,r_L\right)$ denote the trajectory of interaction, where each state is observed at the start point of the time window, while each reward is received at the end point of the window. The environment provides states $\boldsymbol{s}_{1:L}$ and rewards $r_{1:L}$ to train the pacing agent. The agent makes sequential selection decisions $a_{1:L}$ based on current policy $\pi$, and interacts with the environment for further improving the utility of policy.

Figure~\ref{overview:rltp} gives an overview of our proposed \textsf{RLTP} framework. The network architecture is based on Dueling DQN~\cite{wang2016dueling}, which approximates state-action value function via two separate deep neural networks. To jointly optimize the two objectives of impression count and delivery performance, \textsf{RLTP} employs tailored reward estimator which considers four key aspects to guide the agent policy learning: satisfying the guaranteed impression count, encouraging smoothness, maximizing the traffic value, and penalizing the over-delivery as well as the drastic change of selection probability. To effectively fuse multiple reward values to a unified scalar, we further parameterize the fusion weights and learns them with network parameters together. 

We describe the basic elements in the MDP, including state representation (Section~\ref{method:state}), action space (Section~\ref{method:action}) and reward estimator (Section~\ref{method:reward}). We then introduce how to approximate the state-action value function with deep neural networks and the optimization of network parameters (Section~\ref{method:optimization}).

\subsection{State Representation: Delivery Status}\label{method:state}
At each time window $T_i$'s start point, the state representation $\boldsymbol{s}_i\in\mathcal S$ reflects the delivery status after windows $\left\{T_1,\ldots,T_{i-1}\right\}$. Specifically, we design the state representation $\boldsymbol{s}_i$ that consists of statistical features, user features, ad features and context features: 
\begin{itemize}
    \item \textbf{Statistical features}: describing delivery status. Feature list is as follows:
        \begin{itemize}
            \item [-] Observed impression count sequence of previous $i-1$ time windows: $\left\{ N_{1},\ldots, N_{i-1} \right\}$
            \item [-] Cumulatively observed impression count of previous windows: $N_{1:i-1}=\sum_{t=1}^{i-1}N_{t}$
            \item [-] Current delivery completion rate: $D_{i-1}=N_{1:i-1}/N_{\text{target}}$
            \item [-] Delivery completion rate sequence of previous $i-1$ time windows: $\left\{D_1,\ldots,D_{i-1}\right\}$
            \item [-] The difference of completion rates between two adjacent windows: $\Delta_{i-1} = \left(N_{1:i-1} - N_{1:i-2}\right)/N_{\text{target}}$
            \item [-] Averaged CTR of previous $i-1$ time windows: $\overline{CTR}_{1:i-1}=\sum_{t=1}^{i-1}C_{t} / \sum_{t=1}^{i-1}N_{t}$, where $C_*$ denotes click count.
            \item [-] CTR sequence of previous windows: $\left\{{CTR}_{1},\ldots,{CTR}_{i-1}\right\}$
            \item [-] Averaged selection probability of previous $i-1$ time windows: $\overline{a}_{1:i-1}=\frac{1}{i-1}\sum_{t=1}^{i-1}a_t$
        \end{itemize}
    \item \textbf{Context features}: describing context information of traffic requests, including decision time (the values of week, hour and minute).      
    \item \textbf{User and Ad features}: describing user profile and ad information. They contribute to maximize the selected traffic value, but also make state space extremely large. In practice, we use \textit{pretrained} user and ad embeddings and they are \textit{not updated} during agent learning. See Appendix~\ref{supp:agent} for details. We consider window $T_{i-1}$'s averaged user/ad embedding.  
\end{itemize}

We perform discretization to continuous features and transform their values to IDs. Thus for both discrete and continuous features, we represent them as embeddings.\footnote{For sequence feature, we represent it as the mean-pooling of its elements' embeddings. }  The state representation $\boldsymbol{s}_i$ is the concatenation of all features' embeddings.

\subsection{Action Space}\label{method:action}

At the time window $T_i$, given the state representation $\bm s_i$, the pacing agent need to produce a selection probability $a_i\in[0, 1]$ as an action for all traffic requests in this time window. If the total number of requests in window $T_i$ is $N_\mathrm{req}^{i}$, there will be $a_i\cdot N_\mathrm{req}^{i}$ requests that are selected by the pacing agent to fill-in ads.

For ease of handling, we choose discrete action space for producing selection probability at each decision time. Specifically, the action space of selection probability $a_i$ is defined with a step size of 0.02: $\mathcal A = \{0, 0.02, 0.04, \ldots, 0.98, 1.0\}$. In practice we found that the above discrete space is enough to obtain ideal pacing performance. 

Note that in experiments, we compare the discrete action space with continuous action space, see Section~\ref{exp:mainresults} for  analysis.

\subsection{Tailored Reward Estimator}\label{method:reward}
The core of our \textsf{RLTP} framework is the reward estimator, which guides the learning of the pacing agent under the delayed impression situation. At the time window $T_i$'s start point, the agent receives a state representation $\bm s_i$ and takes an action $a_i$ produced by its current policy. At $T_i$'s end point (and also $T_{i+1}$'s start point), the agent receives a reward $r_i$ that reflects the delivery completeness, smoothness and performance, and the state transforms to $\bm s_{i+1}$. 

We design tailored reward estimator to jointly optimize the two objectives of impression count and delivery performance. Formally, we decompose the reward to four terms: 1) satisfying the guaranteed impression count, 2) penalizing the over-delivery, 3) penalizing the drastic change of selection probability, and 4) maximizing the value of selected traffic. 

\subsubsection{\textbf{Satisfying Guaranteed Impression Count}} 
The first term aims to encourage that the actual impression count is as closer as possible to the desired value $N_{\text{target}}$. At the end point of window $T_i$, let $N_{1:i}$ be the cumulatively observed impression count of previous windows, and we set the reward term as:
\begin{equation}
    r_i^{(1)} = \left\{
    \begin{aligned}
        \ \exp\left(N_{1:i} / N_\mathrm{target}\right) &,\quad &\text{if } N_{1:i}\leq N_\mathrm{target}+\epsilon\\
        \ 0&,\quad &\text{otherwise}
    \end{aligned}
    \right.
\end{equation} 
In this way, if the final impression count is less than the demand $N_\mathrm{target}$, the pacing agent will still receive positive reward because over-delivery is not happened.

\subsubsection{\textbf{Penalizing Over-Delivery}}
The second term aims to avoid over-delivery (i.e., $N_\mathrm{imp}> N_\mathrm{target}+\epsilon$). At the end point of window $T_i$, we set the second reward term as:
\begin{equation}
    r_i^{(2)} = \left\{
    \begin{aligned}
        \ 1-\exp\left(N_{1:i} / N_\mathrm{target}\right)^2&,\quad &\text{if } N_{1:i}>N_\mathrm{target}+\epsilon\\
        \ 0&,\quad &\text{otherwise}
    \end{aligned}
    \right.
\end{equation}
If the observed impression count $N_{1:i}$ is larger than $N_\mathrm{target}+\epsilon$, this term returns a large negative value to illustrate that the agent's previous decisions are fail to guarantee the impression count demand and result in over-delivery. 

\subsubsection{\textbf{Penalizing Drastic Change of Selection Probability}}
The third term is for delivery smoothness to reach active audiences at different time periods. 
We measure the difference of observed impression counts between adjacent two windows to produce:
\begin{equation}
    r_i^{(3)} = \left\{
    \begin{aligned}
        \ +1 &,\quad &\text{if } \left| \frac{N_{1:i}-N_{1:i-1}}{N_{1:i-1}} \right| < \text{c} \\
        \ 0&,\quad &\text{otherwise}
    \end{aligned}
    \right.
\end{equation}
where $\text{c}>0$ is a small constant. If the difference is less than $\text{c}$, this term returns a positive value, and the pacing agent is encouraged to smoothly select traffic during the whole delivery period. 

\subsubsection{\textbf{Maximizing Selected Traffic Value}}
The fourth term aims to encourage that the pacing agent can bring good delivery performance. At the end point of window $T_i$, let $C_{1:i}$ be the cumulatively click count of previous windows, and we set the reward term as:
\begin{equation}
    r_i^{(4)} = \exp\left(100\cdot\left(\frac{C_{1:i}}{N_{1:i}}-CTR_{\text{base}}\right)\right)
\end{equation}
where $CTR_{\text{base}}$ is a predefined CTR that the policy should beat.

Finally, the reward $r_i$ for each decision step is the weighted sum of the above four terms:
\begin{equation}\label{equation:reward}
    r_i=\eta_1\cdot r_i^{(1)} + \eta_2\cdot r_i^{(2)} + \eta_3\cdot r_i^{(3)} + \eta_4\cdot r_i^{(4)}
\end{equation}
where $\eta_*$ is the weight factor. Next section we will detail how to obtain the factors.

\subsection{Network Architecture and Optimization}\label{method:optimization}
We adopt Dueling DQN~\cite{wang2016dueling} as the network architecture of \textsf{RLTP}, which is a modified version of DQN~\cite{mnih2013playing}. Further, to effective fuse the reward terms, we parameterize the weight factors. 

\subsubsection{\textbf{Approximate Value Functions with Neural Networks}} 
Let $R_i  \stackrel{\text{def}}{=}\sum_{j=i}^L \gamma^{j-i} r_j$ be the discounted cumulative reward of decision step $T_i$. For the pacing policy $\pi$, a Q-function $Q^{\pi}(\boldsymbol{s},a)$ and a value function $V^{\pi}(\boldsymbol{s})$ represent the long-term values of the state-action pair $(\boldsymbol{s},a)$ and the state $\boldsymbol{s}$ respectively:
\begin{equation}
\begin{aligned}
    Q^{\pi}(\boldsymbol{s},a) &= \mathbb E_{\tau} \left[ R_i \mid \boldsymbol{s}_i=\boldsymbol{s}, a_i=a, \pi\right] \\
    V^{\pi}(\boldsymbol{s}) &= \mathbb E_{a\sim \pi}\left[ Q^{\pi}(\boldsymbol{s},a)\right]
\end{aligned}
\end{equation}
The difference of them represents each action's quality, which is defined as the advantage function $A^{\pi}(\boldsymbol{s},a)$:
\begin{equation}
    A^{\pi}(\boldsymbol{s},a) = Q^{\pi}(\boldsymbol{s},a) - V^{\pi}(\boldsymbol{s})
\end{equation}

Due to the large space of state-action pairs, Dueling DQN employs two separate deep neural networks $V_{\Theta}(\cdot)$ and $A_{\Phi}(\cdot)$ to approximate $V^{\pi}(\boldsymbol{s})$ and $ A^{\pi}(\boldsymbol{s},a)$ respectively: 
\begin{equation}
    V_{\Theta}(\boldsymbol{s}) = \mathsf{DNN}(\boldsymbol{s}; \Theta)\,,\quad 
    A_{\Phi}(\boldsymbol{s},a) = \mathsf{DNN}(\boldsymbol{s}, a; \Phi) 
\end{equation}
where $V_{\Theta}(\cdot)$ and $A_{\Phi}(\cdot)$ are a stack of fully-connected layers. $\Theta$ and $\Phi$ denote learnable parameters. State and action are represented as embeddings to be fed in the networks. 

The approximated Q-function and the pacing policy are: 
\begin{equation}
\begin{aligned}
     Q_{\Theta,\Phi}(\boldsymbol{s},a) &= V_{\Theta}(\boldsymbol{s}) + \left( A_{\Phi}(\boldsymbol{s},a) - \frac{1}{|\mathcal A|}\sum_{a'\in\mathcal A}A_{\Phi}(\boldsymbol{s},a') \right)\,,  \\
     \pi(\boldsymbol{s}) &= \arg\max_{a'\in\mathcal A}Q_{\Theta,\Phi}(\boldsymbol{s}, a')\,.
\end{aligned}
\end{equation}

\subsubsection{\textbf{Learning Reward Fusion Factors $\eta$}}
The reward estimator contains multiple terms (Equation~\ref{equation:reward}) and each has a factor $\eta_*$. To obtain appropriate factors for reward fusion, we parameterize the factors as a trainable vector $\boldsymbol{\eta}=(\eta_1,\eta_2,\eta_3,\eta_4)^{\top}$, and feed it into the neural networks to learn the factors~\cite{Friedman2018Generalizing}:
\begin{equation}
    V_{\Theta}(\boldsymbol{s},\boldsymbol{\eta}) = \mathsf{DNN}(\boldsymbol{s},\boldsymbol{\eta}; \Theta)\,,\quad 
    A_{\Phi}(\boldsymbol{s},a,\boldsymbol{\eta}) = \mathsf{DNN}(\boldsymbol{s}, a,\boldsymbol{\eta}; \Phi) \,.
\end{equation}

\subsubsection{\textbf{Optimization}}
The pacing agent interacts with the environment to collect trajectories as training data (replay buffer). The pacing policy is optimized via the following objective by gradient descent, which takes a batch of $(\bm s_i, a_i, r_i, \bm s_{i+1})$ sampled from replay buffer at each iteration:
\begin{equation}
    \min_{\Theta, \Phi} \sum_{(\bm s_i, a_i, r_i, \bm s_{i+1})} \left( r_i + \gamma \max_{a'\in\mathcal A}Q_{\Theta, \Phi}(\bm s_{i+1}, a') - Q_{\Theta, \Phi}(\bm s_{i}, a_i)\right)^2\,.
\end{equation}

Note that in real world applications, it is impossible that the pacing policy directly interacts with production system to perform online training. We design a simulator as the environment to perform offline training. See Section~\ref{exp:simulator} for details.

\section{Experiments}\label{sec:exp}
We conduct offline and online experiments on real world datasets, and aim to answer the following research questions:
\begin{itemize}
    \item \textbf{RQ1} (effectiveness): Does our \textsf{RLTP}  outperform competitors algorithms of impression pacing for preloaded ads?
    \item \textbf{RQ2} (impression count demand): Does the learned policy of \textsf{RLTP} satisfy the impression count demand of advertisers, as well as the smoothness requirement during ad delivery? 
    \item \textbf{RQ3} (delivery performance demand): Does the learned policy of \textsf{RLTP} select high-value traffic during ad delivery?
    \item \textbf{RQ4} (online results): Does our \textsf{RLTP} achieve improvements on core metrics in  industrial advertising platform?
\end{itemize}

\subsection{Experimental Setup}
In real world advertising platforms, deploying a random policy online then directly performing interaction with production systems to train the policy is almost impossible, because the optimization of an RL agent need huge numbers of samples collected from the interaction process, and this will hurt user experiences and waste advertisers' budgets. Therefore, we train an offline simulator using the historial log collected from a behavior policy (i.e., the current pacing algorithm of our production system), and the training and offline evaluation processes of \textsf{RLTP} are based on the interaction with the simulator.

\subsubsection{\textbf{Offline Datasets and Simulator}}\label{exp:simulator}
The historial log comes from a one-week delivery process on our advertising platform, which is collected by the pacing algorithm described in Section~\ref{baseline:pid}. The log records the following information:
\begin{enumerate}
    \item timestamps of traffic requests
    \item selection decisions of behavior policy
    \item timestamps of actual impressions
    \item user feedbacks (i.e., click or not)
    \item meta-data of user and ad profiles.
\end{enumerate}

The number of total requests in the historial log is \textit{8.85 million}, and the actual impression count is \textit{1.12 million}. We regard each day as one delivery period, and the pacing policy need to produce sequential selection probabilities at 0:00, 0:05, ..., 23:50 and 23:55, containing \textit{288 decisions}.

We utilize the historial log to build our offline simulator that is used for learning and evaluating the pacing agent. We preprocess this log to the format of quads (state, action, reward, next state), see Section~\ref{method:state},~\ref{method:action} and~\ref{method:reward} for the definitions of such elements. The simulator is built via learning a state transition model $\hat{\mathcal P}(\boldsymbol{s}, a)\rightarrow \boldsymbol{s}'$ and a reward model $\hat{\mathcal R}(\boldsymbol{s}, a)\rightarrow r$. Based on the preprocessed historial log, we can obtain a large size of triplets $(\boldsymbol{s}_i, a_i, \boldsymbol{s}_{i+1})$ and $(\boldsymbol{s}_i, a_i, r_{i})$ and then use them to train models for simulation.

In practice, recall that the statistical features of state representation (see Section~\ref{method:state}) and the final form of reward (see Section~\ref{method:reward}) are the combination from observed impression count $N_*$ and click count $C_*$, thus we simplify the simulator learning via estimating them in state $\boldsymbol{s}_{i+1}$ given those in $\boldsymbol{s}_i$ and action $a_i$, The model architecture is a neural network, and it has multi-head outputs for producing different features (i.e., observed impression count and click count). The mean squared error loss function is used to optimize continuous output. See Appendix~\ref{supp:simulator} for more details.

\subsubsection{\textbf{Evaluation Metrics}}
For offline experiments, we measure different aspects of the pacing agent to evaluate the learned policy:
\begin{itemize}
    \item \textbf{Delivery completion rate}: the proportion of actual impression count to the desired impression count after the whole delivery period, which should be close to 100\%.  
    \item \textbf{Click-through rate (CTR)}: the ratio of click count to actual impression count after the whole delivery period. It reflects the traffic value and delivery performance. 
    \item \textbf{Cumulative reward}: it directly measures the final utility of the learned pacing agent. 
\end{itemize}

\subsubsection{\textbf{Implementation Details}}
In offline experiments, we set the desired impression count $N_\mathrm{target}$ as \textit{0.15 million}, and $\epsilon$ is 10\% of it. We duplicately run  3 times and report the averaged metric. 

See Appendix~\ref{supp:agent} for more details.

\begin{table}[t]
\footnotesize
\centering
\caption{Offline experimental results of competitors.}
\begin{tabular}{lccc}
\toprule
\textbf{Methods} & \textbf{Completion Rate} & \textbf{CTR} & \textbf{Cumulative Reward} \\ 
\midrule
\textsf{Stats. Rule-based Adjust.} & 151.8\% & 7.327\%  & 3984.5\\
\textsf{Pred.-then-Adjust.} & 120.3\% & 7.353\% & 5637.3  \\
\textsf{RLTP-continuous}  & 116.5\% & 7.560\% & 5750.1 \\
\textsf{RLTP}  & \textbf{107.7\%} & \textbf{7.621\%} & \textbf{6080.2} \\
\bottomrule
\end{tabular}
\label{table:rq1}
\end{table}

\subsection{RQ1: Comparative Results}
\subsubsection{\textbf{Comparative Algorithms}}
We compare the \textsf{RLTP} to well-designed baselines for preloaded ads impression pacing. In traditional pacing algorithms, the feedback signals used for controlling need to be \textit{immediate}, while in our problem the impression signals can only be observed at the end of the delivery period due to the preloading strategy. Thus, existing pacing algorithms cannot be directly adopted to tackle delayed impression without modification.   

We design two baselines to impression pacing for preloaded ads. 
\begin{itemize}
    \item \textbf{\textsf{Statistical Rule-based Adjusting}}. Based on historial delivery logs, we statistically summarize the mapping between observed impression count and actual count for each time window. This mapping is utilized for next delivery: At each time window, we use the observed count to lookup the mapping that transforms it to actual count (but it may not be accurate). If it is near to the desired count, we directly set the selection probability as zero to avoid over-delivery.
    \item \textbf{\textsf{Prediction-then-Adjusting}}, modified from the state-of-the-art framework~\cite{cheng2022adaptive} (Section~\ref{baseline:pid}). It is a two-stage solution that contains an actual impression count prediction model and a PID controller. The first stage estimates potential delayed count as immediate feedback signals for the controlling of the second stage. 
\end{itemize}
\textsf{Prediction-then-Adjusting} is the current policy of our production system and now serving the main traffic, thus it is a strong baseline. 

To compare continuous and discrete action spaces, we also introduce a variant of \textsf{RLTP} as a baseline:
\begin{itemize}
    \item \textbf{\textsf{RLTP-continuous}}. We modify the network architecture from the Dueling DQN to proximal policy optimization (PPO)~\cite{schulman2017proximal} for adapting continuous action space, which learns to produce selection probability directly using a policy function $\pi(a\mid \boldsymbol{s})$. Note that learning factors cannot be applied to it. 
\end{itemize}

\begin{figure}[t]
\centering
\centerline{\includegraphics[width=0.46\columnwidth]{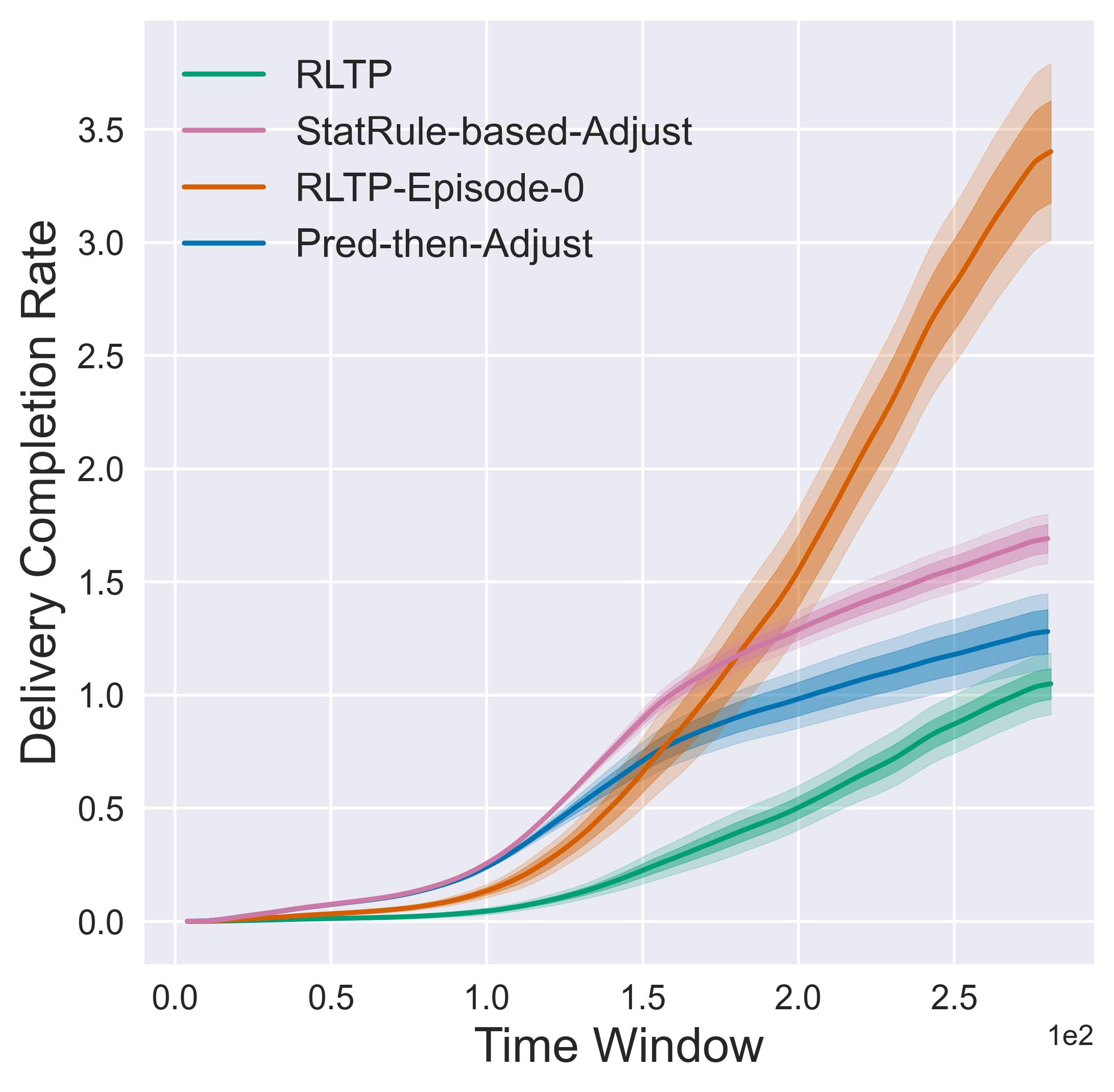}}
\vspace{-1em}
\caption{For each competitor, the trend of delivery completion rate during all 288 time windows in the delivery period.}
\label{exp:baselines}
\vspace{-1.5em}
\end{figure}

\begin{figure}[t]
\centering
\subfloat[Cumulative reward.]{
    \centering
    \includegraphics[width=0.45\columnwidth]{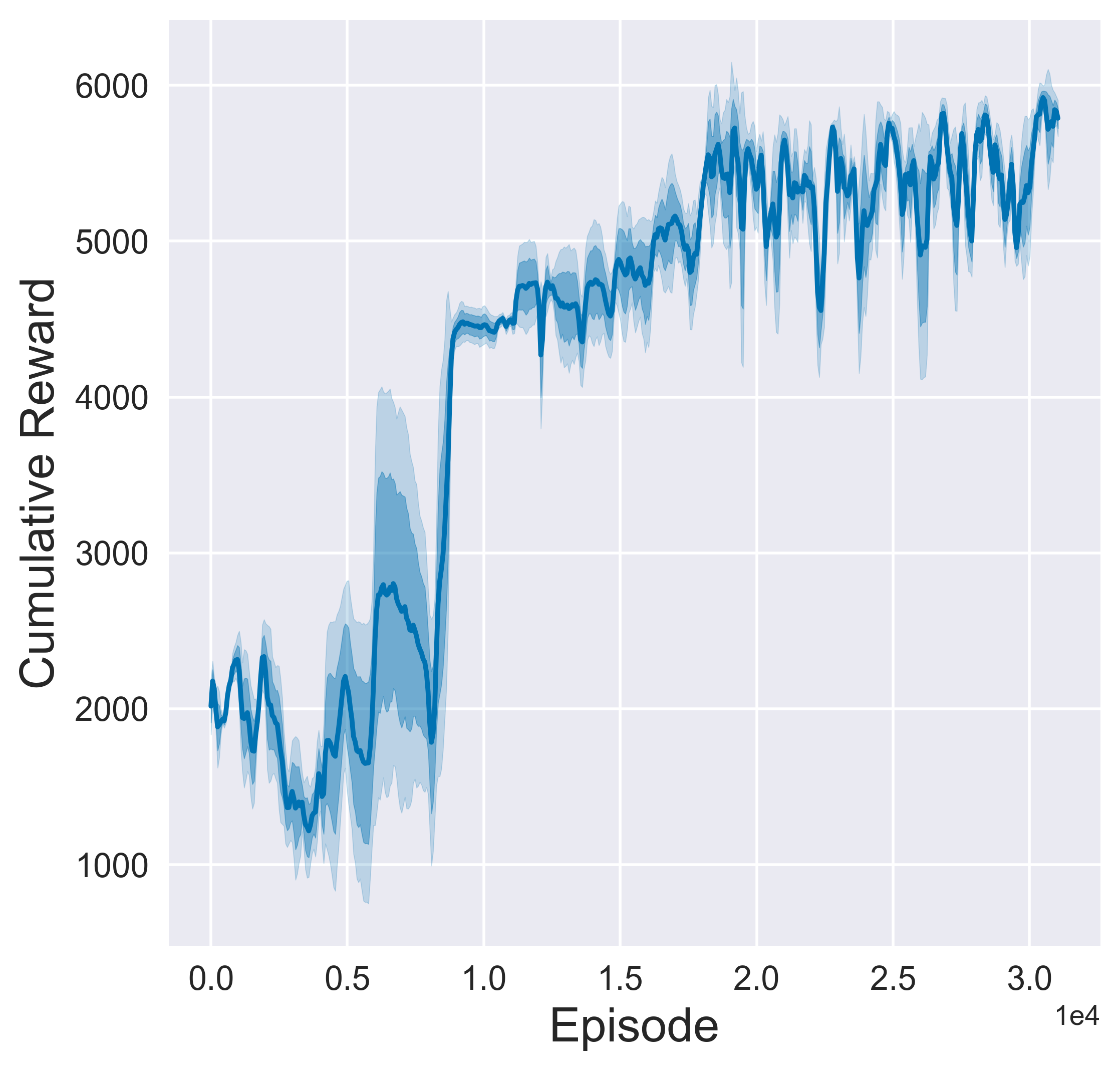}
}
\subfloat[Delivery completion rate.]{
    \centering
    \includegraphics[width=0.44\columnwidth]{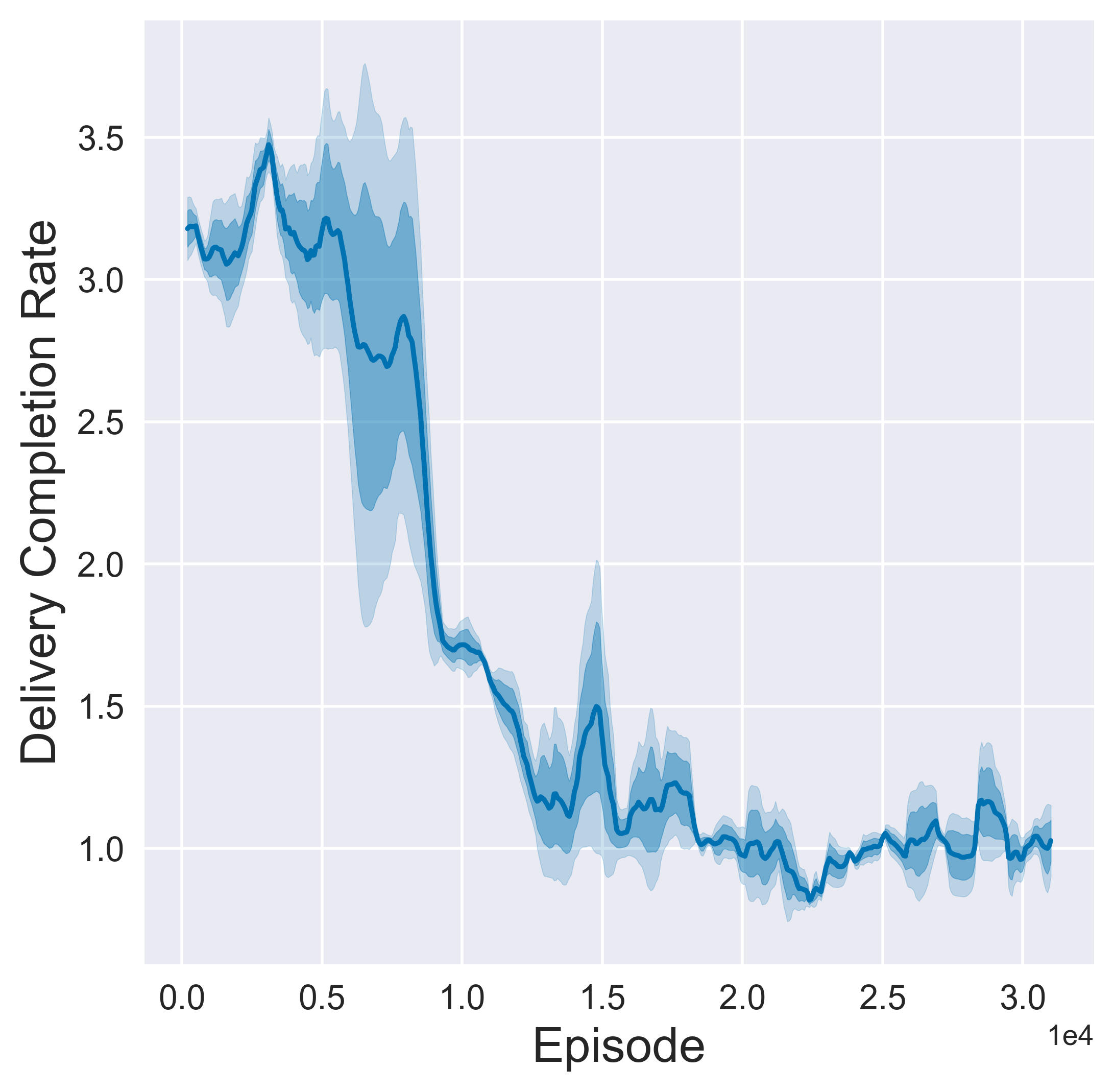}
}
\vspace{-1em}
\caption{The curves of cumulative reward and delivery completion rate during training (30,000 episodes).}
\label{exp:impressioncount}
\vspace{-1.5em}
\end{figure}

\subsubsection{\textbf{Comparative Results}}
Table~\ref{table:rq1} shows the results of comparative algorithms, where the two baselines' performances are also evaluated by the simulator. We observe that \textsf{RLTP} achieves the best results on all three metrics, verifying its effectiveness to preloaded ads pacing. The main limitation of two baselines is that both of them rely on statistics from historical delivery procedures. This results in that their timeliness is problematic when the traffic distribution has changed compared to historical statistics, restricting their pacing performances. Although the \textsf{Prediction-then-Adjusting} algorithm considers real-time responses in the first stage's prediction, it still employs a prior expected impression count as the goal for controlling. In contrast, the pacing agent of \textsf{RLTP} is aware of real-time delivery status via well-designed state representation, avoiding over-delivery and selecting high-value traffic effectively.

For each of the two baselines and \textsf{RLTP}, Figure~\ref{exp:baselines} illustrates the trend of completion rate at all 288 time windows in the whole delivery period during evaluation. Compared to \textsf{RLTP}, the \textsf{Stat. Rule-based Adjust.} algorithm results in serious over-delivery, and the completion rate of \textsf{Pred.-then-Adjust.} is also not reasonable.

Figure~\ref{exp:impressioncount} shows the curves of cumulative reward and completion rate during \textsf{RLTP}'s training process. At the beginning of training, the pacing agent brings serious over-delivery (i.e., the completion rate is larger than 300\%). As the increasing of training episode, the cumulative reward of the agent continuously gains and the delivery rate tends to coverage and finally becomes around 100\%. This demonstrates that the reward estimator effectively guides the agent policy learning to meet the demand of impression count.

\subsubsection{\textbf{Compare Continuous and Discrete Action Spaces}}\label{exp:mainresults}
From Table~\ref{table:rq1}, we observe that the \textsf{RLTP} based on discrete action space performs slightly better than the \textsf{RLTP-continuous}. Furthermore, compared to \textsf{RLTP-continuous}, we observe that \textsf{RLTP} has  following advantages: 1) The number of training episodes is far less than that of \textsf{RLTP-continuous}, thus its training efficiency is better, and 2) the consistency between policy's selection probability and environment's traffic CTR is also more reasonable. Based on the above reasons, we choose \textsf{RLTP} based on discrete action space as the final version. Detailed illustrations are shown in Appendix~\ref{supp:actionspace}.

\subsection{RQ2: Analysis on Impression Count and Smoothness Demands}
We show how the learned pacing agent of \textsf{RLTP} avoids over-delivery to guarantee the impression count demand. We then analyze how it performs on smoothness during the whole delivery period. 

\subsubsection{\textbf{Avoiding Over-Delivery}}
\begin{figure}[t]
\centering
\centerline{\includegraphics[width=0.48\columnwidth]{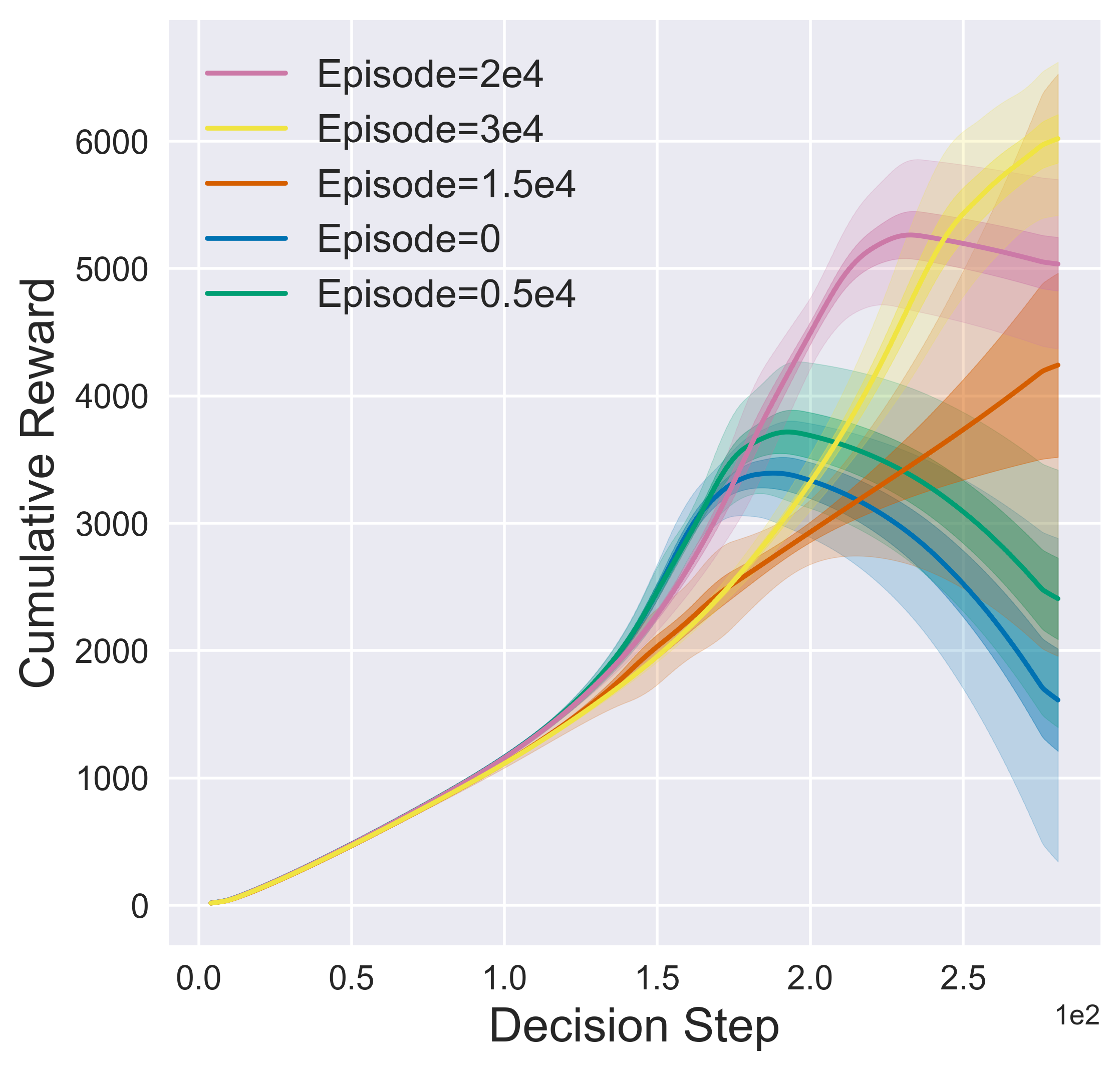}}
\vspace{-0.5em}
\caption{At different training episodes, the trends of cumulative reward at all 288 decision steps.}
\label{exp:overdelivery}
\vspace{-1em}
\end{figure}

Figure~\ref{exp:overdelivery} shows the trend of cumulative reward within an episode (288 decision steps). We observe that, at early episodes, the cumulative reward drops drastically after 100 decision steps, which means that there exists over-delivery (the reward term $r^{(2)}$ brings a large negative value). At later episodes, the cumulative reward always grows with decision step, verifying that the pacing agent gradually learns ideal policy that avoids over-delivery via the trial-and-error interaction with environment.

\subsubsection{\textbf{Delivery Smoothness}}
According to the slope rate of each trend line in Figure~\ref{exp:baselines}, we observe that our \textsf{RLTP}'s pacing agent makes most of ad impressions evenly distributed to 288 time windows and thus performs smooth delivery. However the two baselines do not guarantee the smoothness well. This shows that the learned policy of \textsf{RLTP} can reach active audiences at diverse time periods to improve ads popularity.

\begin{figure}[t]
\centering
\subfloat[The trend of CTR.]{
    \centering
    \includegraphics[width=0.46\columnwidth]{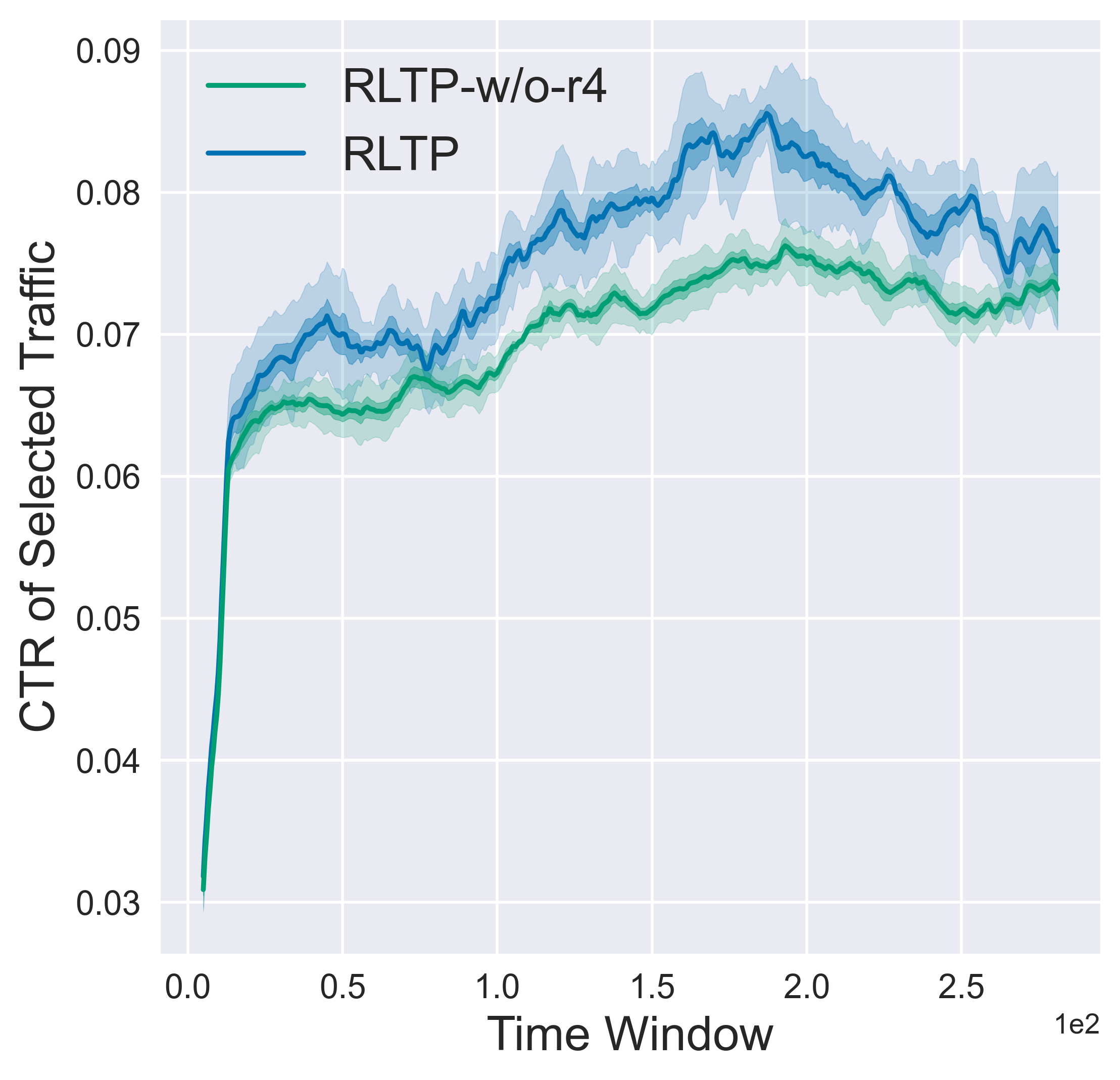}
}
\subfloat[The consistency of selection probability w.r.t. traffic CTR of historical log.]{
    \centering
    \includegraphics[width=0.53\columnwidth]{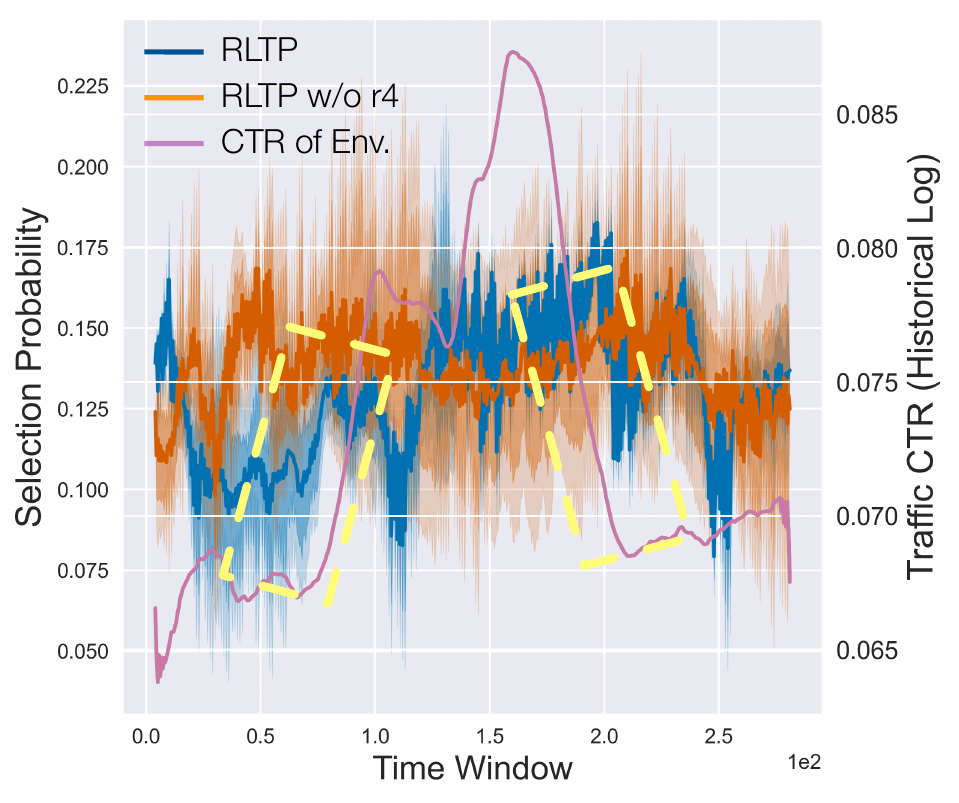}
}
\caption{Effects of maximizing traffic value.}
\label{exp:deliveryperformance}
\vspace{-1em}
\end{figure}

\subsection{RQ3: Analysis on Delivery Performance}
In this section, we turn to analyze how the pacing agent of \textsf{RLTP} achieves good delivery performance. 

\subsubsection{\textbf{Effect of Traffic Value Modeling}}
To verify the effect of maximizing selected traffic value during  policy, Figure~\ref{exp:deliveryperformance} (a) compares the CTR of selected traffic by two pacing agents: the one is our agent that trained using our reward estimator, and the other's training process does not consider the fourth reward term $r^{(4)}$ that maximizes the selected traffic value. We observe that without modeling traffic value during policy learning, the agent misses potential high-value traffic. This verifies that explicitly incorporating traffic value reward is effective to help improve delivery performance.

\subsubsection{\textbf{Correlation between Policy's Selection Probability and Environment's Traffic CTR}} 
We analyze whether the learned pacing agent possesses the ability of dynamically adjusting selection probability based on environment's traffic value. Specifically, we first compute the \textit{traffic CTR} of different time windows according to the \textit{historial log}, which reflects the environment's traffic value trend. We expect that if a time window's traffic CTR is higher, our pacing can increase the selection probability to obtain high-value traffic as much as possible and improve delivery performance.  

Figure~\ref{exp:deliveryperformance} (b) compares the trends of agent's selection probability and environment's traffic CTR. We can see that the consistency of \textsf{RLTP}'s agent and environment is better than the agent that does not consider the fourth reward term. Concretely, we highlight several decision period segments to illustrate the consistency differences of the two agents. This demonstrates that \textsf{RLTP}'s agent is sensitive to environment's traffic CTR and can dynamically adjust selection probability to achieve higher delivery performance.

\subsection{RQ4: Online Experiments}
In this section, we first introduce how to deploy the \textsf{RLTP} framework online to our brand advertising platform, and then report the online experimental results in production traffic. 

\subsubsection{\textbf{Online Deployment}}
The online architecture of \textsf{RLTP} for impression pacing generally contains two parts. One is the pacing agent learning module that performs policy training, and the other is the online decision module that produces selection probability in real-time for ad delivery. 

\textbf{I.} For the pacing agent learning module, the offline training procedure is introduced in Section~\ref{sec:method}, which returns a learned policy $\pi$ via the interaction process with the simulator. To deploy the pacing policy to production system, the key is that it should be adapt to new traffic distribution and user interests. Therefore, the pacing agent learning module is running in a cyclical fashion, in which {fresh} ad delivery logs are utilized to {incrementally update} the simulator model, and the update cycle is typically one day. After the simulator model is updated, we restore the pacing policy $\pi$ and  incrementally update its learnable parameters via small numbers of (e.g., hundreds of) episodes with the updated simulator. 

\textbf{II.} For the online decision module, it first loads the update-to-date pacing policy that consists of state embedding tables and value functions, and also accesses the desired impression count $N_{\text{target}}$ at the start time of the current delivery period (typically at 0:00). Then it cyclically executes the following adjusting steps for sequentially producing selection probabilities (typically 5-minutes per cycle):
\begin{description}
    \item[Step1] Calculate the observed impression count from the start time to now. If it has been larger than the target count or the current time window is the last one, set the selection probability to zero immediately; otherwise, turn to Step 2.
    \item[Step2] Collect the statistical features and context features (see Section~\ref{method:state}) in real-time, and lookup averaged user and ad embeddings from database. Then feed these features to the loaded pacing policy's state embedding tables, and finally obtain the state representation $\boldsymbol{s}$.
    \item[Step3] Feed the state representation to the loaded pacing policy's value functions $V_{\Theta}(\cdot)$ and $A_{\Phi}(\cdot)$, and produce a set of action values $\left\{Q(\boldsymbol s, a)\right\}_{a\in\mathcal A}$. Then choose the action with the highest value as the selection probability $a$.
    \item[Step4] For all traffic requests sent by online publisher at this time window, select and fill-in our ad with probability $a$. If a request is selected, send a key-value pair of (unique identification, impression=0) to our impression counter. The impression counter is in charge of constantly listening whether a filled ad is displayed on the publisher, and sets impression=1 after display. Then turn to Step 1.
\end{description}

\begin{figure*}[t]
\centering
\centerline{\includegraphics[width=2.05\columnwidth]{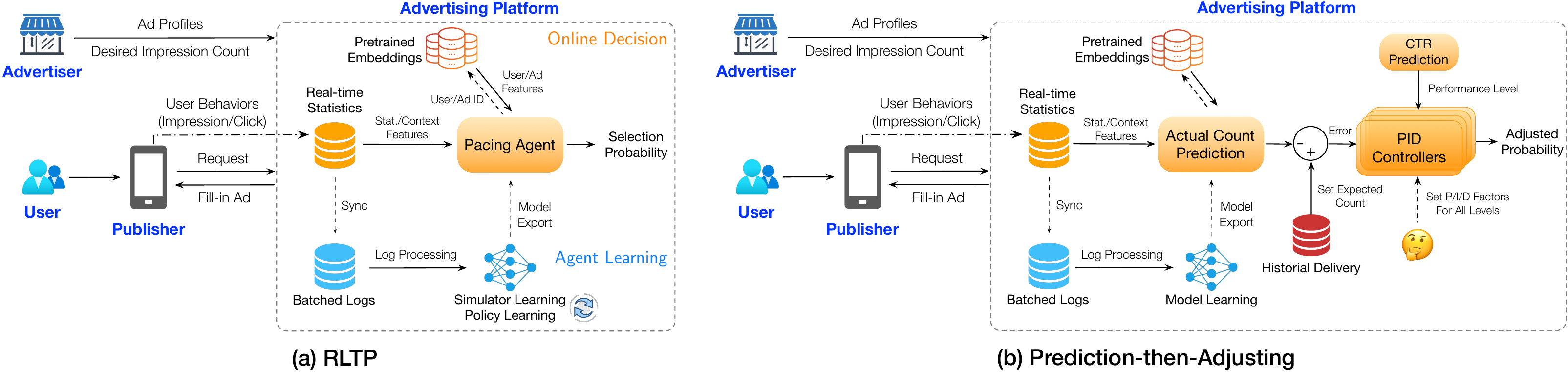}}
\vspace{-1em}
\caption{Online deployment for \textsf{RLTP} and \textsf{Prediction-then-Adjusting}.}
\label{fig:onlinearch}
\vspace{-1em}
\end{figure*}

Figure~\ref{fig:onlinearch} shows the comparison of online deployment for \textsf{RLTP} and \textsf{Prediction-then-Adjusting}. Compared to the two-stage solution, \textsf{RLTP} simplifies the online architecture to a unified agent. More importantly, unlike the \textsf{Prediction-then-Adjusting} which needs estimates actual impressions and sets ``hard'' expected impression counts based on historical delivery procedures to guide the controlling for current delivery, \textsf{RLTP} does not rely on such estimation and setting. Instead, the pacing policy is learned to maximize long-term rewards via trail-and-error interaction process with the simulator, which performs relatively ``soft'' controlling. For both algorithm design and system architecture, our \textsf{RLTP} framework takes a next step to the problem of impression pacing for preloaded ads.

\subsubsection{\textbf{Online A/B Test}}
We conduct online A/B test for one week on our brand advertising system. The base bucket employs the two-stage algorithm \textsf{Prediction-then-Adjusting} (the strongest algorithm for our production system currently), and the test bucket employs the pacing policy learned by our \textsf{RLTP} framework. The online evaluation metrics contain delivery completion rate and CTR, measuring whether the pacing algorithm guarantees impression count and delivery performance demands respectively.

The online A/B test shows that the delivery completion rates of the base and test buckets are 118.4\% and \textbf{111.6\%} respectively,  and the test bucket brings the relative improvement of \textbf{+2.6\%} on CTR. The results demonstrate that \textsf{RLTP} brings significant uplift on core online metrics and satisfies advertisers' demands better.

\section{Related Work}
Generally, the problem of impression pacing for preloaded ads is related to three research fields. The first is pacing control algorithms~\cite{abrams2007optimal,borgs2007dynamics,bhalgat2012online,agarwal2014budget,xu2015smart,geyik2020impression}. Existing studies focus on budget pacing that aims to spend advertiser budget smoothly over time. Representative  approaches~\cite{agarwal2014budget,xu2015smart} adjust the probability of participating in an auction to control the budget spending speed, and they rely on immediate feedbacks (e.g., win/lose a bid) for effective adjusting. 

The second is the allocation problem of guaranteed display advertising~\cite{Bharadwaj2012shale,chen2012hwm,hojjat2014delivering,zhang2017efficient,fang2019large,zhang2020request,cheng2022adaptive}. These work formulates a matching problem for maximizing the number of matched user requests (supply nodes) and ad contracts (demand nodes) on a bipartite graph, where the delivery probabilities are computed offline and then the delivery speed is controlled online by pacing algorithms. Although the impression count demand is considered in the matching problem, these studies can not tackle the delayed impression issue.

The third is delayed feedback modeling~\cite{chapelle2014modeling,ktena2019addressing,yasui2020feedback,yang2021capturing}. Existing work focuses on modeling users' delayed behavior, such as purchase. Such delay nature is caused by user inherent decision, other than online publishers. Thus the delayed pattern is essentially different to our problem that the delayed situation is mainly controlled by online publishers and also affected by delayed requests from users. 

Overall, current studies do not touch the impression pacing problem under preloading strategy from online publishers. Our work makes the first attempt and tackles this problem with a reinforcement learning framework.

\section{Conclusion}
We focus on a new research problem of impression pacing for preloaded ads, which is very challenging for advertising platforms due to the delayed impression phenomenon. To jointly optimize both impression count and delivery performance objectives, we propose a reinforcement learning to pace framework \textsf{RLTP}, which employs tailored reward estimator to satisfy impression count, penalize over-delivery and maximize traffic value. Experiments show that \textsf{RLTP} consistently outperforms compared algorithms. It has been deployed online in our advertising platform, achieving significant uplift to delivery completion rate and CTR.

\bibliographystyle{ACM-Reference-Format}
\bibliography{src-shortversion.bib}

\appendix
\section*{Supplementary Material}
We provide more implementation details and computation resource for reproducibility in Section~\ref{supp:implenentation}. We also show the comparison between \textsf{RLTP-continuous} and \textsf{RLTP} in Section~\ref{supp:actionspace} to justify why we choose discrete action space. 

\section{Implementation Details}\label{supp:implenentation}

\subsection{Simulator Learning}\label{supp:simulator}
The offline data is the historial log collected from a one-week delivery process, containing \textit{8.85 million} requests and \textit{1.12 million} impressions. The function of the offline simulator is to provide next state and four reward terms given current state and sampled action. Recall that all statistical features of state and reward terms of reward estimator are the combination  of  the observed impression count and click count. Therefore, in practice we simplify the simulator learning via estimating the two values, and both next state and reward can be computed via the estimated values. 

Specifically, we organize the historial log to the form of (\textsf{inputs1}, \textsf{inputs2}, \textsf{outputs}) based on adjacent time windows: (\textsf{inputs1} = $i$-th window's observed impression/click counts and user/ad features, \textsf{inputs2} = $i$-th window's selection probability, \textsf{outputs} = $(i+1)$-th window's observed impression/click counts). The simulation model is a deep neural network that estimates \textsf{outputs} given \textsf{inputs1} and \textsf{inputs2}. Note that user/ad features are retrieved from pretrained embeddings and are not updated. 

After training, the simulation model  estimates next windows impression/click counts to produce next state and four reward terms, thus it can be used to perform interaction with pacing agent.

\subsection{Policy Learning}\label{supp:agent}
In our offline experiments, we set the desired impression count $N_\mathrm{target}$ as \textit{0.15 million} for one delivery period, and $\epsilon$ is 10\% of it. 

In \textsf{RLTP}, the user and ad features in state representations are represented using pretrained user and ad embeddings, which are not updated during agent learning. Specifically, our user and ad embeddings are produced by a graph neural network model~\cite{hamilton2017inductive}, where the node set contains users and ads, and user-ad click behaviors are regarded as edges. The learned embeddings are stored in database and can be retrieved by the pacing agent. Both the value function $V_{\Theta}(\cdot)$ and the advantage function $A_{\Phi}(\cdot)$ are three-layered neural networks, with output sizes of [200, 100, 1] and [200, 100, 50] respectively. Each state/context feature used for state representation is represented as a 4-dim embedding, and each candidate action id $a\in\mathcal A$ also uses 4-dim embedding. The base CTR in the fourth reward term is set to the averaged CTR of historial log. The discount factor $\gamma$ in culumative reward formulation is set to 0.99. 

We train the pacing agent for around 30,000 episodes, and duplicately run  3 times to report the averaged metric in the Section~\ref{sec:exp}.

\subsection{Computation Resource}\label{supp:resource}
Both the simulator learning and the policy learning procedures are done on  8 Tesla V100 GPUs. 

\section{Detailed Comparison of Discrete and Continuous Action Spaces}\label{supp:actionspace}
In Section~\ref{exp:mainresults}, we breifly introduce the advantages of \textsf{RLTP} (which employs discrete action space) compared to \textsf{RLTP-continuous}. Here we give concrete illustrations. 

\begin{figure}[t]
\centering
\subfloat[Delivery completion rate during training (200,000 episodes).]{
    \centering
    \includegraphics[width=0.46\columnwidth]{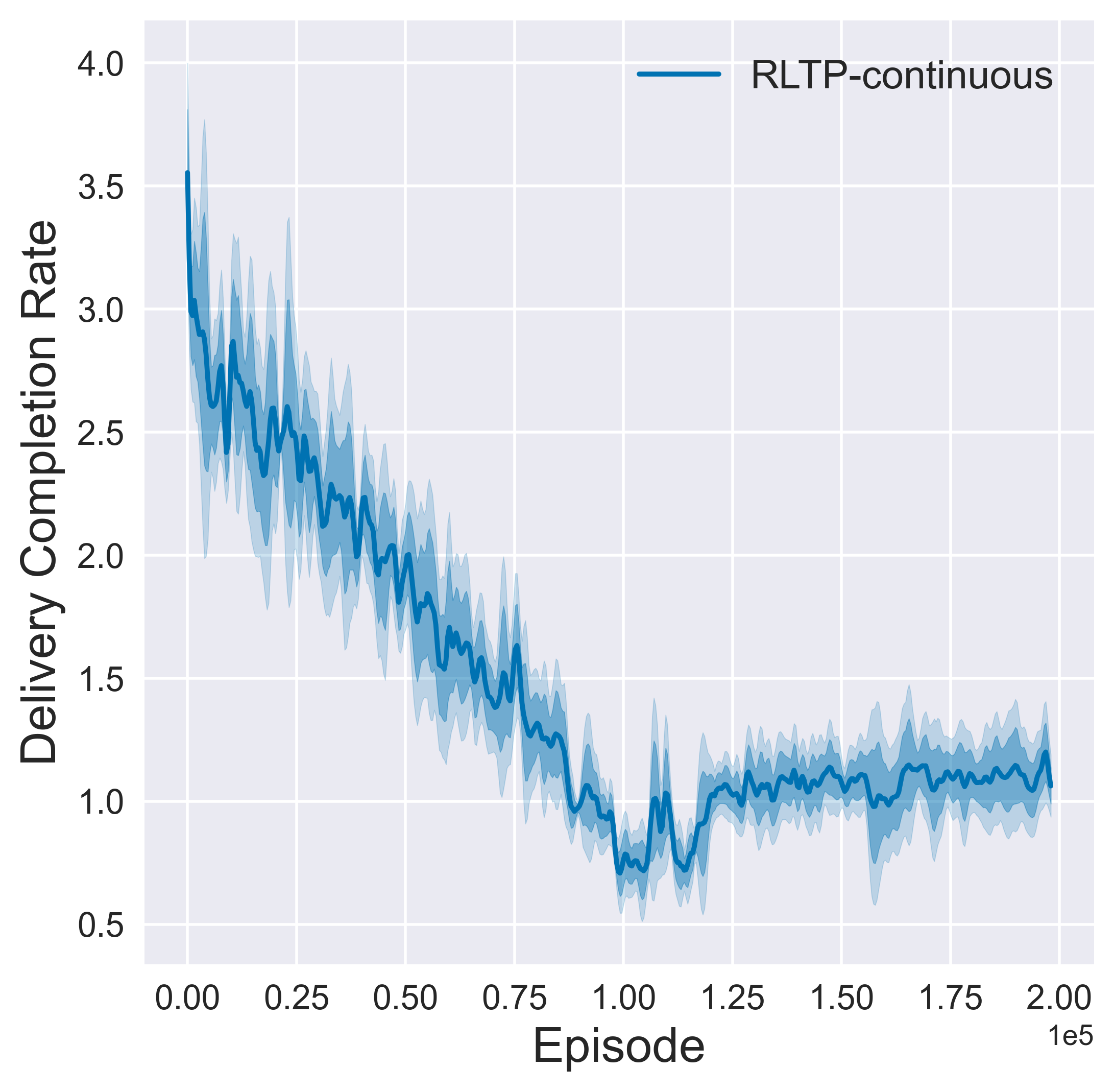}
}
\subfloat[Selection probability at all 288  steps.]{
    \centering
    \includegraphics[width=0.53\columnwidth]{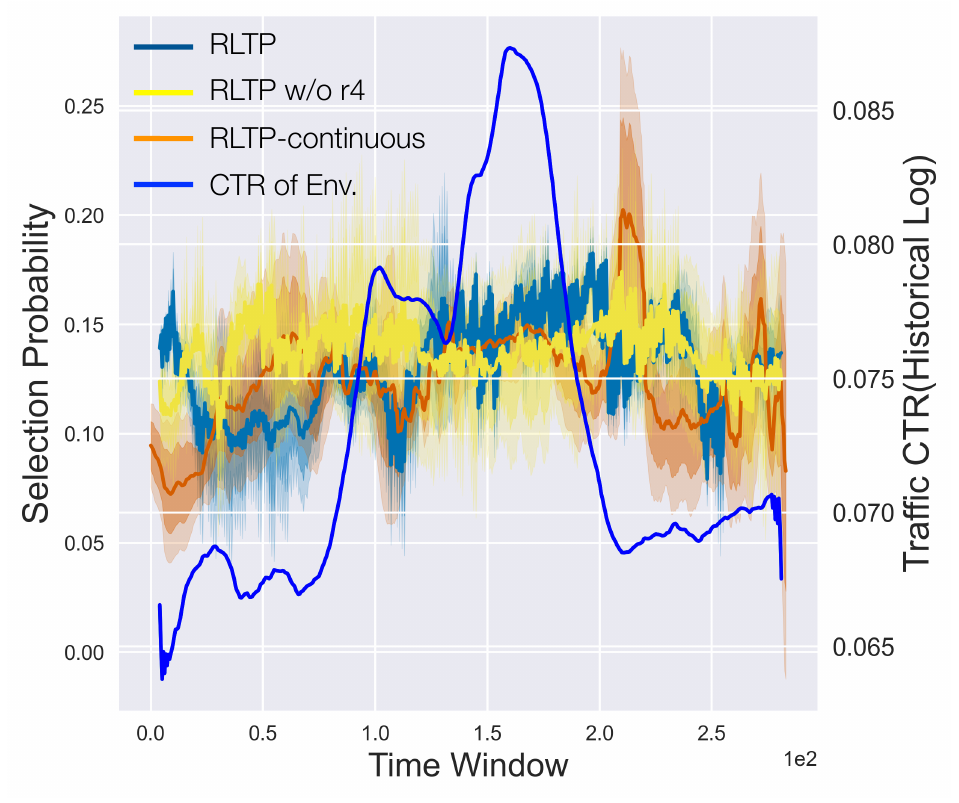}
}
\caption{Training efficiency and traffic value of \textsf{RLTP-continuous}.}
\label{exp:ppo}
\vspace{-1em}
\end{figure}

First, we compare the training efficiency. We observe that the \textsf{RLTP-continuous} needs around 200,000 episodes to achieve ideal pacing policy as shown in Figure~\ref{exp:ppo} (a), while \textsf{RLTP} only needs around 30,000 episodes as in Figure~\ref{exp:impressioncount}. In industrial applications, training efficiency is a key factor for production system, thus we prefer \textsf{RLTP} for online deployment. 

Second, we show the delivery performance. From Table~\ref{table:rq1} we can see that CTR metric of \textsf{RLTP} and \textsf{RLTP-continuous} is similar. We further investigate the trend of selection probability. Figure~\ref{exp:ppo} (b) illustrates the trend of selection probability of \textsf{RLTP-continuous}'s pacing agent, and we observe that the consistency between selection probability and environment's traffic CTR is not as good as \textsf{RLTP} shown in Figure~\ref{exp:deliveryperformance} (b). 

The above results demonstrate that \textsf{RLTP} shows advantages compared to \textsf{RLTP-continuous}. Therefore we choose \textsf{RLTP} with discrete action space as the final version, and in future work we shall explore how to improve \textsf{RLTP-continuous} for production system.

\end{document}